\documentclass[aip, jap, reprint]{revtex4-1}
\usepackage[colorlinks,bookmarksopen]{hyperref}
\usepackage{graphicx}
\usepackage{dsfont} 
\usepackage{xspace}
\usepackage{amsmath, amssymb}
\usepackage{threeparttable}
\usepackage{tabularx}
\usepackage[english]{babel} 
\usepackage{color}

\graphicspath{{/}}

\newcommand{\m}[1]{#1}

\newcommand{\quotes}[1]{``#1''}
\newcommand{\f}{\mathbf}                         
\newcommand{\bra}[1]{\left\langle #1 \right|}     % bra-Vectors
\newcommand{\ket}[1]{\left| #1 \right\rangle}     % ket-Vectors

\newcommand{\matrixelem}[3]{\left\langle #1 \right| #2 \left| #3
\right\rangle} 
\newcommand{\etal}{\textit{et al.}\xspace}

\newcommand{\eg}{\mbox{e.\,g.}\xspace}
\newcommand{\eV}{\mbox{e\hspace{0.08mm}V}\xspace}          
\newcommand{\meV}{\mbox{me\hspace{0.08mm}V}\xspace}        
\newcommand{\phpl}{\phantom{+}}

\begin{document}

% Use the \preprint command to place your local institutional report number 
% on the title page in preprint mode.
% Multiple \preprint commands are allowed.
%\preprint{}

\title{Structure-Related Optical Fingerprints in the
       Absorption Spectra of Colloidal Quantum Dots:
       Random Alloy vs.~Core/Shell Systems
       }
%Title of paper

% repeat the \author .. \affiliation  etc. as needed
% \email, \thanks, \homepage, \altaffiliation all apply to the current author.
% Explanatory text should go in the []'s, 
% actual e-mail address or url should go in the {}'s for \email and \homepage.
% Please use the appropriate macro for the type of information

% \affiliation command applies to all authors since the last \affiliation
%command. 
% The \affiliation command should follow the other information.

\author{Daniel Mourad}
\email{dmourad@itp.uni-bremen.de}
\affiliation{Institute for Theoretical Physics, University of Bremen,
Otto-Hahn-Allee 1, 28359 Bremen, Germany}

% Collaboration name, if desired (requires use of superscriptaddress option in
%\documentclass). 
% \noaffiliation is required (may also be used with the \author command).
%\collaboration{}
%\noaffiliation

\date{\today}


\begin{abstract}
We argue that the experimentally easily accessible optical 
absorption spectrum
can often be used to distinguish between a random alloy phase
and a stoichiometrically equivalent core/shell realization of ensembles of 
 monodisperse
colloidal semiconductor quantum dots without the need for more advanced
structural characterization tools. 
Our proof-of-concept is performed by
conceptually straightforward
exact-disorder tight-binding calculations. 
The underlying stochastical tight-binding
scheme only parametrizes bulk band structure properties and does not employ
additional free parameters to calculate the optical absorption spectrum, which is
an easily accessible experimental property.  The method is applied to selected
realizations of type-I Cd(Se,S)
 and type-II (Zn,Cd)(Se,S) alloyed quantum dots with an underlying zincblende 
 crystal structure and the corresponding core/shell
 counterparts.
\end{abstract}

% insert suggested PACS numbers in braces on next line
%\pacs{71.15.Ap, 71.20.Nr, 71.23.-k, 73.40.Kp, 78.55.Cr}

\maketitle %\maketitle must follow title, authors, abstract and \pacs

%%%%%%%%%%%%%%%%%%%%%%%%%%%%%%%%%%%%%%%%%%%%%%%%%%%%%%%%%%%%%%%%%%%%%%%%%%%%%

%%%%%%%%%%%%%%%%%%%%%%%%%%%%%%%%%%%%%%%%%%%%%%%%%%%%%%%%%%%%%%%%%%%%%%%%%%%%%
\section{Introduction\label{sec:introduction}}
%%%%%%%%%%%%%%%%%%%%%%%%%%%%%%%%%%%%%%%%%%%%%%%%%%%%%%%%%%%%%%%%%%%%%%%%%%%%%

It is well established that the optoelectronic properties of colloidal semiconductor
quantum dots (QDs, also known as semiconductor nanocrystals) can be tuned by
means of alloying, i.e., the use of solid solutions of a multitude of compounds.
These alloyed systems exhibit a usually complicated and nonlinear
combination of the material properties of their constituents. As an example, 
it is possible to vary the stoichiometry of few-nanometer
sized (Cd,Zn)(Se,S) alloyed QDs to shift the optical gap over the 
visible spectral range.~\cite{deng_band_2009} To make the distinction between a randomly alloyed QD and
the corresponding stochimetrically equivalent core/shell counterpart is difficult,
especially for small diameters of the nanocrystals. 
It has been shown that rather sophisticated
characterization tools like Raman spectroscopy can be used to confirm the presence
of solid-solution nanocrystals,~\cite{aubert_homogeneously_2013} as well
as partial alloying.~\cite{tschirner_interfacial_2011} However, an identification based on an optical
measurement (such as the UV-vis absorption spectrum) would clearly be 
advantageous due to the simplicity of the approach. Also, it would circumvent the
need for a modification  and re-interpretation 
of the already rather involved bulk Raman selection rules for nanosized
systems with complicated boundary conditions.

In a recent publication,\cite{mourad_random-alloying_2014} we have shown 
for the example of Cd(Se,S)
that alloyed semiconductor QDs show fingerprints in their absorption
spectrum that can be traced back to the inherent breakdown of point group 
symmetries in random systems, which subsequently leads to a lifting of selection rules.
We further pointed out that 
these features can be reproduced and analyzed by means of conceptionally 
simple stochastic tight-binding (TB) simulations that emulate the disorder degree
of freedom on finite ensembles. Moreover, we concluded that a feasible analysis
is already possible on a single-particle level including the light-matter coupling
in the dipole approximation.

In the present work, we will explore whether similar calculations can  be 
used in general to unambiguously identify randomly alloyed QDs---as opposed to their
stoichiometrically identical but phase-separated core/shell 
counterparts---by their linear absorption spectrum. The method is applied to
type-I Cd(Se,S) and type-II (Zn,Cd)(Se,S) alloyed quantum dots; the latter will be 
compared to core/shell realizations with a CdS core and a ZnSe shell and vice versa. 
We restrict our discussion to QDs with a diameter of $d \approx3$ nm, which
corresponds to 5 conventional lattice constants of the zincblende crystal structure
for the materials under consideration.

%%%%%%%%%%%%%%%%%%%%%%%%%%%%%%%%%%%%%%%%%%%%%%%%%%%%%%%%%%%%%%%%%%%%%%%%%%%%%
\section{Theory\label{sec:theory}}
%%%%%%%%%%%%%%%%%%%%%%%%%%%%%%%%%%%%%%%%%%%%%%%%%%%%%%%%%%%%%%%%%%%%%%%%%%%%%

The underlying theoretical framework of the here employed stochastic TB
approach has already been explained in detail in
Ref.\,\onlinecite{mourad_random-alloying_2014} and the references therein. Therefore,
we will here only give a brief outline of the most relevant features and
will put more emphasis on eventual modifications and simplifications which are
required to cope with the numerical complexity of the task at hand.

\subsection{Tight-Binding Model}
The empirical bulk TB approach employed in this paper is based on a parametrization
of eight Wannier-like orbitals $\ket{\f{R} \alpha} \otimes \ket{\sigma}$, where
$\alpha \in \{s,p_x,p_y,p_z\}, \sigma = \{\uparrow, \downarrow\}$ and $\f{R}$ labels 
the underlying fcc lattice sites of the zincblende structure.
\cite{loehr_improved_1994} In this approach, non-vanishing tight-binding matrix elements
(TBME) of the bulk Hamiltonian  
$\matrixelem{\f{R} \alpha}{H_\text{bulk}}{\f{R}' \alpha'}$ up to the second nearest 
neighbor shell are used to accurately reproduce the dressed electronic band structure
of one conduction
and three valence bands per spin projection throughout the whole Brillouin zone in 
accordance with input from experiment or \textit{ab initio} quasiparticle  calculations.
The 
corresponding input parameters for CdS, CdSe and  ZnSe as well as their
original sources can be found in Table \ref{tab:materialParameters}. 

%%%%%%%%%%%%%%%%%%%%
\begin{table*}
\caption{Material parameters that were used to fit the tight-binding
matrix elements, including the references. CB and VB stand for conduction and
valence band, respectively.}
\label{tab:materialParameters}
\begin{center}
\begin{tabular}{lll|>{$}l<{$}r|>{$}l<{$}r|>{$}l<{$}r}
{} & {} & {} & \text{CdSe} & Ref. & \text{CdS} & Ref. &\text{ZnSe} & Ref.\\

\hline
\hline
$E_\text{g}$ & bulk band gap & (\eV) & \phpl 1.68 &\citenum{adachi_handbook_2004}
                      & \phpl 2.46 &\citenum{adachi_handbook_2004}
                      & \phpl 2.68 &\citenum{blachnik_numerical_1999}\\
$m_\text{c}$ & CB effective mass &($m_0$)\textsuperscript{*} & \phpl 0.12  &\citenum{kim_optical_1994}
                                          & \phpl 0.14 &\citenum{adachi_handbook_2004}
                                          & \phpl 0.147 &\citenum{kim_optical_1994}\\
$\gamma_1$ & VB Luttinger parameters & & \phpl 3.33  &\citenum{kim_optical_1994}
              & \phpl 4.11 &\citenum{adachi_handbook_2004}
              & \phpl 2.45 &\citenum{holscher_investigation_1985}\\
$\gamma_2$ &{}& {} & \phpl 1.11  & \citenum{kim_optical_1994}
              & \phpl 0.77 &\citenum{adachi_handbook_2004}
              & \phpl 0.61 &\citenum{holscher_investigation_1985}\\
$\gamma_3$ &{}& {} & \phpl 1.45  & \citenum{kim_optical_1994}
              & \phpl 1.53 &\citenum{adachi_handbook_2004}
              & \phpl 1.11 &\citenum{holscher_investigation_1985}\\
$\Delta_\text{so}$ & spin-orbit splitting & (\eV) & \phpl 0.41 &\citenum{kim_optical_1994}
                           & \phpl 0.078 &\citenum{adachi_handbook_2004}
                           & \phpl 0.43 &\citenum{kim_optical_1994}\\
$X_1^\text{c}$ & CB X point energy & (\eV) &\phpl 2.94 & \citenum{blachnik_numerical_1999}
                       & \phpl 5.24 &\citenum{adachi_handbook_2004}
                       & \phpl 4.41 &\citenum{blachnik_numerical_1999}\\
$X_5^\text{v}$ & VB X point energies & (\eV) & -1.98 & \citenum{blachnik_numerical_1999}
                       & -1.9 &\citenum{adachi_handbook_2004}
                       & -2.08 &\citenum{blachnik_numerical_1999}\\
$X_3^\text{v}$ & {} & (\eV) & -4.28 & \citenum{blachnik_numerical_1999}
                       &-4.8 &\citenum{adachi_handbook_2004}
                       & -5.03 &\citenum{blachnik_numerical_1999}\\
$\Delta E_\text{vb}$ & VB offset & (\eV) &\phpl  0.4  &\citenum{monch_empirical_1996}
                             &\phpl  0 &\citenum{monch_empirical_1996}
                             &\phpl  0.35 &\citenum{monch_empirical_1996}\\
$a$ &conventional lattice constant & (\AA) & \phpl 6.078  & \citenum{kim_optical_1994}
            & \phpl 5.825 &\citenum{adachi_handbook_2004}
            & \phpl 5.668 &\citenum{kim_optical_1994}\\
\hline
\end{tabular}
\newline
\textsuperscript{*} Free electron rest mass.
\end{center}
\end{table*}
%%%%%%%%%%%%%%%%%%%%%%%

\subsection{Simulation of randomly alloyed and core/shell QDs} For the 
randomly alloyed A$_x$B$_{1-x}$ QDs, the site-diagonal TBMEs of the pure constituents are
determined stochastically, where the probability of using an A or B TBME
is given by the respective compositions
$x$ and $1-x$. To cover relevant portions of the visible spectrum,
100 electron and 1000 hole energies $e_{i,n}, h_{i,n}$ and TB wave functions
$\psi^{e/h}_{i,n}$ are obtained
by means of exact diagonalization, where $n$ labels the different microstates
(mimicking the individual QD in a large sample). 
\m{The term \quotes{microstate} is here used as common in statistical physics
and designates a specific microscopically identifiable configuration of an alloyed
QD that is characterized by the exact spatial distribution of the lattice sites 
with either A or B material.}
This procedure is then repeated $N=25$ times
for each stoichiometric ratio, 
which turned out to be a suitable compromise between the convergence accuracy and the
 numerical effort. The energy scales of A and B are aligned
via the bulk valence band offset value $\Delta E_\text{vb}$.
Intersite matrix elements between and A and B are approximated by their arithmetic average. 
To emulate the spatial confinement of the approximately spherical QDs, we set
the hopping probability at the surface to zero. This corresponds to a perfect
surface passivation. Strain and relaxation effects are not included in our TB scheme. This
can heuristically be justified by the rather small lattice mismatch for the material combinations 
under consideration ($<3$\% for CdS/ZnSe, $\approx4$\% for CdS/CdSe). Furthermore,
the proper inclusion of bond-length and bond-angle variations into the bulk-parametrized 
TB calculations is conceptually difficult for the surface/volume ratios 
considered here.
\m{We would have to model the impact
of different boundary conditions, e.g., solid-liquid, solid-solid or solid-air
interfaces which has lead to not only ambiguous but even diametrically
opposed results for the nature of the strain field
in  recent studies on II-VI core/shell systems. A striking example is given by
the conflicting conclusions by \citeauthor{tschirner_interfacial_2011} for CdSe/CdS QDs in 
Ref.\,\onlinecite{tschirner_interfacial_2011} (prediction of tensile strain in shell)
and \citeauthor{han_heavy_2015} in Ref.\,\onlinecite{han_heavy_2015} (prediction of compressive strain in shell). 
These studies also reveal that the nature of the strain field is
heavily dependent on the exact modeling of the boundary conditions at the QD surface.
Therefore, our pragmatic decision to use hard-wall boundary conditions and neglect
strain effects altogether prevents the afore cited unresolved
 ambiguities, but will definitely have
an impact on the quantitative accuracy of our results, which cannot be judged 
rigorously within our present study.  On the plus side,
 the
 computational effort is diminished, as the strain field and the subsequent  corrections to our TBMEs would additionally have to be computed
  for each single realization of 
 the QDs in the ensemble, alongside a careful testing and fitting procedure for the 
 distance scaling
 law of the TBME.}

\subsection{Optical properties} 
The transition probability for converting a photon into an electron-hole pair 
is obtained on the footing of Fermi's golden rule,
\begin{equation}\label{eq:FermiRule}
I(\omega) = \frac{2 \pi}{\hbar} \frac{1}{N}
 \sum\limits_{n=1}^{N}
\sum\limits_{i,f} | \langle \Psi^f_{n} |
H_{D} | \Psi^i_{n} \rangle |^2
\delta(E^i_{n} - E^f_{n} - \hbar \omega),
\end{equation}
where $n$ labels the initial/final states $|{\Psi^{i/f}_{n}}\rangle$
of the microscopically distinct configurations with energies $E^{i/f}_{n}$, 
$\hbar$ is the reduced Planck constant, $\omega$ the absorption 
frequency and $H_D$ is the light-matter coupling Hamiltonian in the long-wavelength
approximation. 
\m{In other words, for each fixed fixed stoichiometric 
ratio we first calculate separate $I(\omega)$ curves for each of the
$N=25$ different random QDs and then superpose them in an attempt to
 emulate the spectrum 
of the macroscopic ensemble. 
This also encompasses the model assumption that the indvidual QDs in our 
\textit{in silico} computational \quotes{sample} do not interact explicitly 
by any means.}

As we use hard-wall boundary conditions on finite supercells~\cite{gu_relation_2013} 
and a strictly local 
TB Hamiltonian,\cite{chadi_spin-orbit_1977} the matrix elements (MEs) of $H_D$ can  be
unambigously calculated from the MEs
of the position operator between the electron and hole 
states $|\psi^{e/h}_i \rangle$ which constitute the $|{\Psi^{i/f}_{n}}\rangle$,
\begin{equation}\label{eq:dipoleME}
\mathbf{d}^{eh}_{ij}=e_0\langle\psi^{e}_i|\f{r}|\psi^{h}_j \rangle,
\end{equation}
where $e_0$ is the electron charge, $\f{r}$ is approximated by the Bravais
lattice operator \mbox{$\f{r} \approx \sum_{\f{R} \alpha} 
\ket{\f{R}\alpha}\f{R} \bra{\f{R}\alpha}$}\cite{zak_quantum_1981} and
the $|\psi^{e/h}_i \rangle$ are constituted by their expansion coefficients 
in the TB basis orbitals.
\m{This diagonal form of the position operator is gauge invariant, 
see, \eg, Ref.\,\onlinecite{boykin_dielectric_2001}.}
Albeit the short-range contributions from the wave functions
do not enter explicitly, it should be stressed that our approach is on a 
higher
level of sophistication than the usual multiband cell-envelope
factorization\cite{gu_relation_2013} in the envelope function approximation.
In our model, the
potential landscape of the alloy is resolved in lattice resolution
and the orbital character of the underlying orbitals is also probed.

\subsection{Coulomb effects and dielectric confinement}
Surface polarization effects due to the dielectric mismatch at the QD boundaries
are currently neglected;
the $|{\Psi^{i/f}_{n}}\rangle$ can then in principle be obtained in a basis
of Slater determinants that are constructed from the $|\psi^{e/h}_i \rangle$
after the corresponding Coulomb interaction MEs have been computed
at a desired level of sophistication
(see, e.g.,\,Refs.\,\onlinecite{sheng_multiband_2005, schulz_multiband_2009,
mourad_multiband_2010}
for a detailed description of a TB-based approach).
This procedure (which represents
a simplified TB treatment of the Bethe-Salpeter equation for excitons
\cite{sham_many-particle_1966})
was feasible in Ref.\,\onlinecite{mourad_random-alloying_2014},
where the necessary number of basis states was significantly lower due to the 
focus on a small energy window. Also,
the influence of the additional disorder degree of freedom on the 
Coulomb ME had yet to be examined in detail in this paper. 
However, for a cardinality of the basis
as employed in the present work, such calculations are still not viable
even with the help of modern supercomputers, as it would involve
the calculation of $\sim 10^{11}$ Coulomb MEs per composition.

Consistent with the findings of Delerue \etal in Ref.\,\onlinecite{delerue_excitonic_2000},
we will instead  make use of a cancellation effect and calculate the excitonic 
absorption spectrum directly in the single-particle basis (i.e., dressed
by bulk interactions) without 
the explicit incorporation of Coulomb MEs. 
Their work concentrated on the nature of the fundamental gap of Si nanocrystals,
where they expressed the excitonic gap $\epsilon^\text{exc}_g$
(the energetically lowest transition energy) as:
\begin{equation}\label{eq:excitonicGap}
\epsilon^\text{exc}_g = \epsilon^\text{qp}_g - E^\text{coul} = 
 \epsilon^\text{0}_g + \delta \Sigma - E^\text{coul}
\end{equation}
Here, the quantity
$\epsilon^\text{qp}_g$ is the 
quasiparticle gap and $E^\text{coul}$
is the attractive interaction between a quasi-electron and a quasi-hole.
\mbox{$\delta \Sigma := \epsilon^\text{qp}_g - \epsilon^\text{0}_g$}
is the self-energy correction to the ``independent particle gap''
$\epsilon^\text{0}_g$. The latter is basically defined by
the partitioning of Eq.\,(\ref{eq:excitonicGap}) and can, \eg, be obtained
from an \textit{ab initio} LDA calculation as the HOMO-LUMO difference. 
Within an empirical TB model, augmented by $GW$ and Bethe-Salpeter calculations, 
Delerue \etal then
showed that the QD-specific deviation of the self energy from its bulk value
$\delta \Sigma - \delta \Sigma_\text{bulk}$ actually cancels with
$E^\text{coul}$ for a large range of QD diameters $d \geq 1.5$\,nm.
Furthermore, they showed that the largest contribution to
$\delta \Sigma - \delta \Sigma_\text{bulk}$
can be interpreted in terms of classical electrostatics as a surface-induced
polarization term
$\Sigma_\text{pol} \approx \delta \Sigma - \delta \Sigma_\text{bulk}$
which is not included in a standard non self-consistent scheme like our
present TB model.
As a consequence of the cancelling effect, one is left with 
\begin{equation}
\epsilon^\text{exc}_g \approx  
 \epsilon^\text{0}_g + \delta \Sigma_\text{bulk} = e_1 - h_1,
\end{equation}
where $e_1/h_1$ are the lowest/highest TB electron/hole energies of the QD system.
The 
last equation is valid because $\delta \Sigma_\text{bulk} = 0$ holds in 
an empirical TB approach, where the dressed single-particle gap values 
will effectively
contain all bulk-related interactions by means of the fitting procedure. Also,
they found that the results were transferable to other materials (Ge and C
in their study)
and do not change upon the exact treatment of the surface passivation.
It should, however, be pointed out that their results were obtained for QDs that were
embedded in media with low polarizabilty (e.g.\,vacuum or air), where
the dielectric constant $\varepsilon_\text{out}$ of the medium is close to unity, 
while the Si QDs had a several times larger average
polarizability, reflected in  a corresponding value $\varepsilon_\text{in}$.

\subsection{Generality of approach}

In Appendix \ref{sec:Appendix}, we elaborate why we also consider this approach
sufficiently accurate for the present systems under consideration for a 
\m{discussion}
of the optical spectrum as presented in this paper. Nevertheless, it should
be 
emphasized that the the main results of the present work are of a \emph{qualitative}
nature and do \emph{not} rely on the full
transferability of the canceling effect reported by Delerue \etal to our II-VI  
systems. \m{We also stress again at this point that our calculations do not 
use microscopically realistic surface boundary conditions and neglect strain-induced effects, which
prevents model-intrinsic ambiguities but
hampers quantitatively accurate predictions. However,} as long as the weaker assumptions listed below are valid (as in detail 
discussed in the Appendix), we
are left with $\delta \Sigma - E^\text{coul}\approx \text{const.}$, resulting
only in a  phase independent constant shift of \emph{both}
the core/shell and random alloy 
spectra:
\begin{itemize}
\item The macroscopic dielectric response of the core/shell
and randomly alloyed QDs can be described by a \emph{common}
$\varepsilon_\text{in}^{-1}$. Consequently, 
the surface-induced polarization $\Sigma_\text{pol}$ does not depend on the
phase of the QD, as it is a function of
$\varepsilon_\text{out}^{-1} -
\varepsilon_\text{in}^{-1}$.\cite{lannoo_screening_1995, jackson_classical_1998}
\item The variation of electron-hole 
Coulomb interaction MEs upon phase segregation is 
small on the \eV energy scale that covers the visible range.
\end{itemize}
Both of these points should actually hold for a broad class of materials:
(i) The inverse macroscopic dielectric constants
of QDs consisting of different miscible
isovalent compound semiconductors are usually very similar for small
systems. According to Delerue \etal,~\cite{delerue_concept_2003} the reduction of 
the screening constant
in small crystallites can then mainly be attributed to the breaking 
of the polarizable bonds at the surface.
(ii) The Coulomb interaction MEs are dominated by the strong confinement of the 
charge carrier wave functions. Our
previous work~\cite{mourad_multiband_2010, mourad_random-alloying_2014} has 
shown that dipole selection rules 
react sensitively to alloy-induced disorder, while the electron-hole 
attraction is not overly sensible to such perturbations. This will also
apply to small isovalent core/shell QDs with thin shells, as the confinement 
of the charge carriers
is dominated by the large (in our model infinite)
discontinuity at the QD boundary and the wave functions leak
across the internal interface.

%
%\end{enumerate}

%%%%%%%%%%%%%%%%%%%%%%%%%%%%%%%%%%%%%%%%%%%%%%%%%%%%%%%%%%%%%%%%%%%%%%%%%%%%%
\section{Results and Discussion\label{sec:results}}
%%%%%%%%%%%%%%%%%%%%%%%%%%%%%%%%%%%%%%%%%%%%%%%%%%%%%%%%%%%%%%%%%%%%%%%%%%%%%

%%%%%%%%%%%%%%%%%%%%%%%%%%%%%%%%%
\subsection{Choice of material system and stoichiometric ratios
            \label{subsec:choice}}

\begin{figure}
\centering
\includegraphics[width=\linewidth]{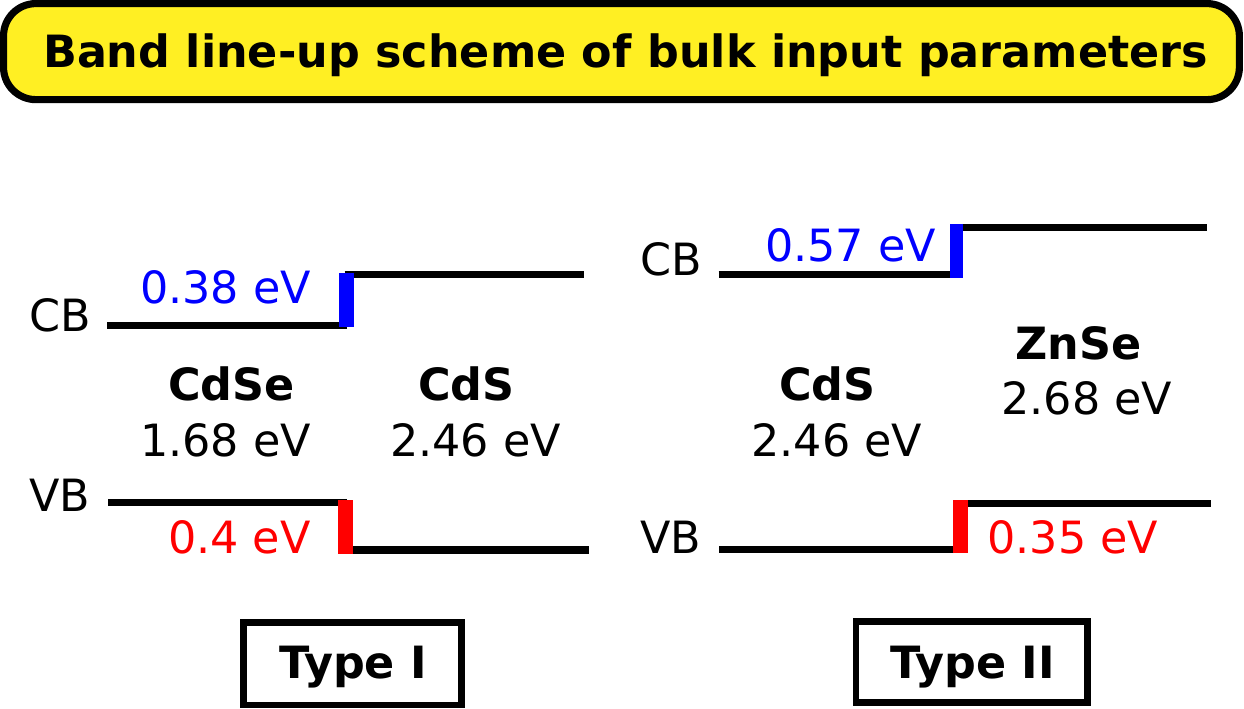}
\caption{
\label{fig:bandLineup}
Scheme of the energetic line-up of the \emph{bulk} conduction band (CB) and  valence band (VB) edges
for the material combinations that are investigated in the present paper, including
the band gap values (black), conduction band offsets (blue) and valence band offsets (red).}
\end{figure}

In the following, we will, for reasons of numerical effort as well as for the sake
of brevity, confine the discussion to selected realizations with representative
stoichiometric ratios. All QDs have a diameter of 5 conventional lattice constants $a$,
which corresponds to $d\approx 3$\,nm for the material systems under consideration.
Throughout this paper, we will employ a name convention that
is derived from the corresponding bulk interface energetics.
 A sketch of the respective bulk 
band line-up (using the material parameters from Table \ref{tab:materialParameters})
can be found in Figure \ref{fig:bandLineup}.

According to Figure \ref{fig:bandLineup}, the CdSe/CdS core/shell system then represents a type-I lineup,
i.e., for large enough systems both the electron and hole states 
in the vicinity of the fundamental gap 
are both predominantly localized in the core region. 
In contrast, 
the type-II CdS/ZnSe
core/shell line-up confines the lowest electron states to the core and the
energetically highest hole states to the shell region for large enough
cores and shells. These features are 
reversed in the type-II ZnSe/CdS system. It should be noted that 
another convention is to 
denote the line-up characteristics based on the ground state
electron and hole wave function
localization, which then introduces intermediate classes 
(e.\,g.,\quotes{type-I-1/2} or \quotes{quasi type-II} where one of the
charges is spread over the QD, while the other is localized to the core or shell,
respectively).~\cite{donega_synthesis_2011} We prefer the more general bulk-related 
name convention 
as our calculations 
expand on a bulk band structure scheme and the carrier overlap naturally enters the 
absorption spectrum by means of the dipole MEs, Eq.\,(\ref{eq:dipoleME}). Moreover,
any of the randomly alloyed QD systems can anyway only be strictly classified by their pure bulk
constituents' line-up.

% Zitat typeIorII \cite{eshet_electronic_2013}

The application-relevant type-I and type-II core/shell realizations usually exhibit thicker
cores and shells than in the present work (see, e.g., Ref.\,\onlinecite{chen_compact_2013}) to enhance
their optical properties. Also, these QDs often exhibit
more or less significant additional modifications like interfacial alloying or a
wurtzite phase.~\cite{chen_compact_2013, bae_controlled_2013} This will not concern us here, as we focus on the distinction
between alloyed QDs and their \quotes{undesired} segregated core/shell
counterparts, rather than on the simulation of tailor-made core/shell QDs.

%\begin{enumerate}
%
\textbf{Type-I QDs.} We simulate CdSe$_{0.25}$S$_{0.75}$ randomly 
alloyed QDs and their CdSe/CdS core/shell
counterpart, which corresponds to  a CdSe core diameter of $5\,a$ coated by a CdS
shell of a thickness of $1\,a$ (hence one layer of CdS unit cells, i.e., approximately
 0.6 nm).
Both of these stoichiometrically equivalent QD types are in principle experimentally 
available\footnote{Some of the forthcoming references deal with wurtzite 
quantum dots, while we assume a zincblende structure.
Similar to the structural analysis in Ref.\,\onlinecite{tschirner_interfacial_2011},
the conclusions of the present work will presumably also hold for wurtzite 
and zincblende/wurtzite polytype QDs because
of the very similar band gaps in the hexagonal phase}
and relevant for a broad range of applications.~\cite{peng_epitaxial_1997, 
swafford_homogeneously_2006,
ouyang_noninjection_2009, brovelli_nano-engineered_2011, chen_compact_2013,
aubert_homogeneously_2013, mourad_random-alloying_2014}
Specifically, the alloyed Cd(Se,S) QDs cover the visible spectrum for the size
simulated here. Therefore, they also 
constitute a suitable benchmark system for the predictive power as well 
as the limitations of our model calculations. 

\textbf{Type-II QDs.} Here, we consider two different classes of 
stoichiometrially equivalent systems:
(a) Cd$_{0.25}$Zn$_{0.75}$S$_{0.25}$Se$_{0.75}$ randomly alloyed QDs and 
their CdS/ZnSe core/shell counterpart with the same geometry as described above.
The former QDs represents a realization of the quarternary 
(Zn,Cd)(Se,S) alloy, as realized, e.g., in Ref.\onlinecite{deng_band_2009}, while the
latter is a promising candidate for QD lasing.~\cite{ivanov_type-ii_2007}
(b)  Cd$_{0.75}$Zn$_{0.25}$S$_{0.75}$Se$_{0.25}$ randomly alloyed QDs and 
their respective ZnSe/CdS core/shell version, which can be used 
in photovoltaics~\cite{ning_solar_2011} or as a starting point for
electronic tuning upon gradual interface alloying.~\cite{boldt_electronic_2014}

% We do not, however, consider the theoretically possible
% corresponding ZnS/Cde and CdSe/ZnS
% core/shell realizations. Cubic ZnS is an insulator with a 
% band gap $E_\text{g}\approx1.6$\,\eV and 
% a valence band offset of $\Delta E_\text{vb} \approx -0.5$\,\eV with 
% respect to CdSe.~\cite{hinuma_band_2014}
%
%\end{enumerate}
%

%%%%%%%%%%%%%%%%%%%%%%%%%%%%%%%%%
%\subsection{Light polarization and line broadening}

In all calculations, we use a light polarization along the [100] direction without loss
of generality.%\footnote{
For measurements on a real macroscopic ensemble, 
where all orientations are present,
any directional dependence will cancel out. In our model calculations for alloys on finite
ensembles, the
statistical averaging over the realizations will approximately have the same effect, 
whereas in the 
core/shell systems the linear optical susceptibility is a tensor of rank two, of which 
the matrix representation has three degenerate eigenvalues in systems with cubic symmetry.
Inversion asymmetry and fine structure related effects are beyond the scope of
the present work and will not contribute on the scale of
energetic resolution that is discussed here.
%}

All calculations use a broadening of the $\delta$-peaks in Eq.\,\ref{eq:FermiRule}
that would correspond to a full width at half maximum of ca.\,120 \meV for a
hypothetical isolated excitation peak at energy $e_i-h_j, i,j = \text{const}$.
This exact value 
had already been used for the detailed analysis in
Ref.\,\onlinecite{mourad_random-alloying_2014}.
The numerical line broadening will in practice also partially 
 incorporate a multitude of effects beyond the
lifetime of the transition, e.g., 
the finite cardinality of the statistical sample as well as 
any size and shape fluctuations and uncertainties
that could in principle be included explicitly in our model
but would significantly enlarge the numerical effort.
It should be pointed out that the above mentioned broadening parameter
has \emph{not} been varied furtherly to ``fine-tune'' the individual spectra. Doing so,
one could achieve a better agreement with the experimental line shapes. However, we refrain from such an approach,
as it would obscure the predictive power and limitations of our stochastic TB scheme.
In contrast, we emphasize that no additional free parameters beyond the necessary 
fitting to bulk electronic properties have been used in our model. 

In the following, we will discuss the optical absorption spectra
for the type-I system and the type-II systems, respectively.
All spectra have been normalized to a common maximum value, as the quantitative discussion
of relative peak heights is beyond the scope of the present work. Instead, we will focus
on the energy/wavelength dependence of selected features, in particular the
phase dependent spectral position of the energetically lowest absorption peak. 

%%%%%%%%%%%%%%%%%%%%%%%%%%%%%%%%%
\subsection{Results for type-I Cd(Se,S) and CdSe/CdS QDs}

\begin{figure}
\centering
\includegraphics[width=\linewidth]{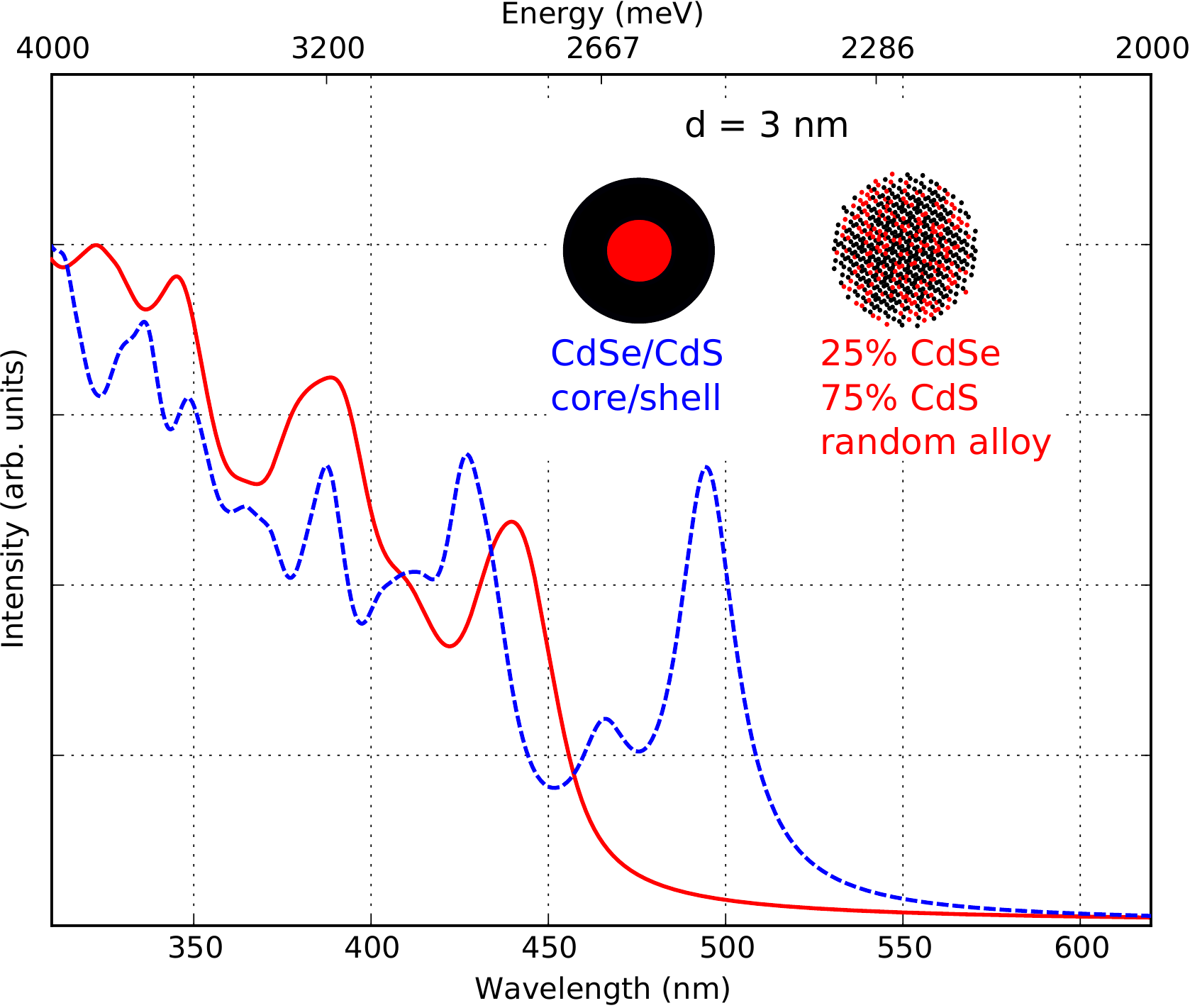}
\caption{
\label{fig:cds_cdse_spectra}
Absorption spectra for CdSe$_{0.25}$S$_{0.75}$ randomly alloyed QDs
(solid red) and their CdSe/CdS core/shell
counterpart (dashed blue).}
\end{figure}

\begin{figure}
\includegraphics[width=\linewidth]{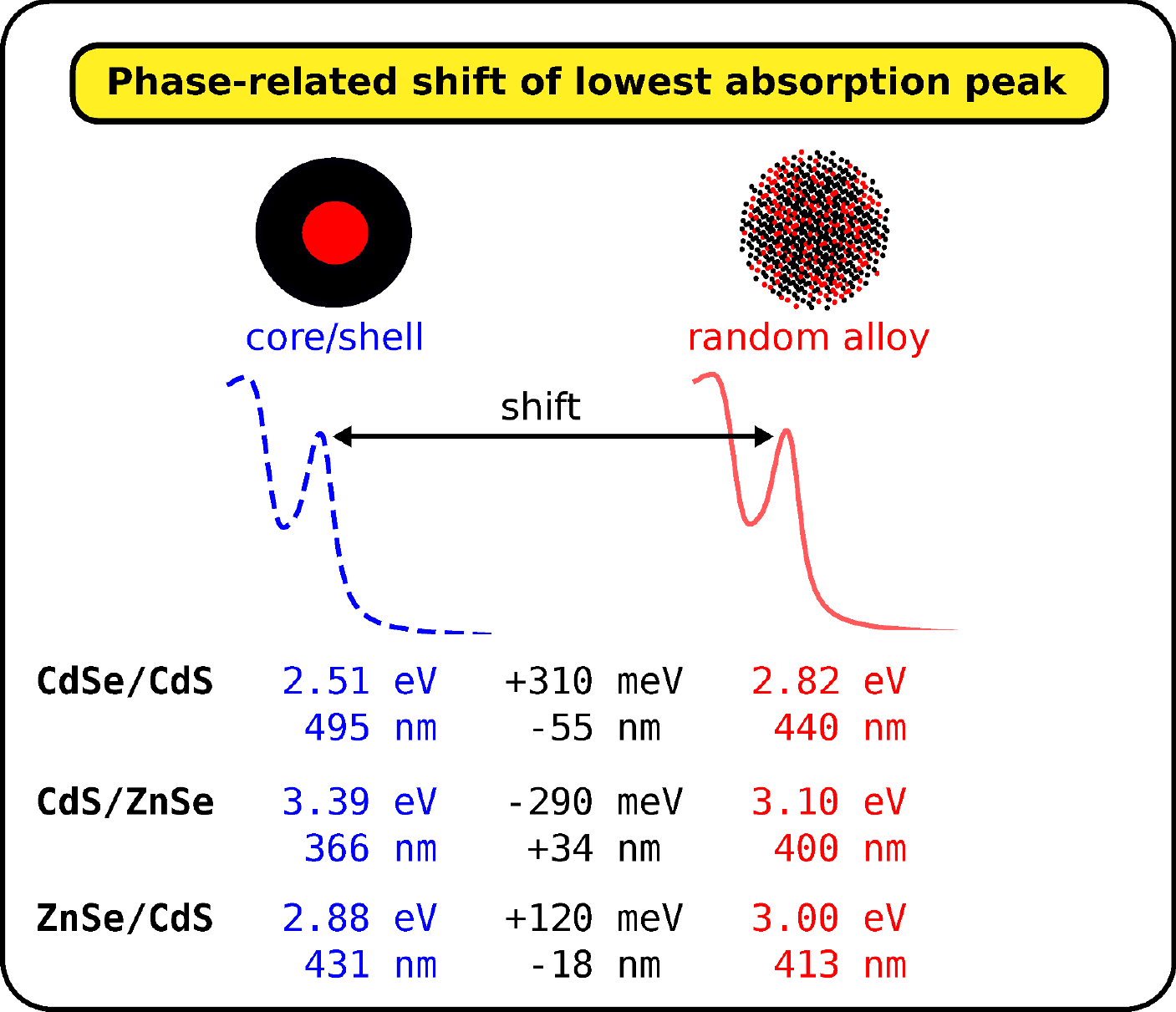}
\caption{
\label{fig:peakShifts}
\centering
Phase-related difference of the energetically lowest absorption peaks for all 
         systems under consideration.}
\end{figure}

\textbf{CdSe$_{0.25}$S$_{0.75}$ vs.~CdSe/CdS.}
The absorption spectra for the CdSe$_{0.25}$S$_{0.75}$ randomly alloyed QDs
and their CdSe/CdS core/shell counterpart can be found in Fig.\,\ref{fig:cds_cdse_spectra}. As depicted in 
Fig.\,\ref{fig:peakShifts}, the energetically lowest 
absorption peak for the alloyed QDs can be found at approximately 2.82 \eV (440 nm). 
This is in very good agreement with the experimental UV-Vis value of 2.88 \eV (431 nm) 
as obtained from the nonlinear bowing parameter for 
the system under consideration.~\cite{mourad_random-alloying_2014} Also,
the randomly alloyed QDs show a small shoulder on the red (smaller energy) side of 
the second main transition peak that can experimentally and theoretically 
be attributed to an optically active
alloying-induced impurity band and vanishes for larger Se contents.~\cite{mourad_random-alloying_2014}
As additional benchmark for the accuracy of our approach, we use the calculated
excitonic gaps for pure CdSe and CdS (absorption curves not depicted here) which are 
obtained as 2.32 \eV (534 nm) and 3.14 \eV (395 nm), respectively. This is in good agreement
(deviation $< 0.1$ \eV) with experimental results from several
studies.~\cite{mourad_random-alloying_2014, bae_controlled_2013}
Upon phase segregation into the core/shell case, we predict a 
redshift of approximately $-310$ \meV ($+55$ nm) for the energetically lowest
transition. 
Also, the calculated core/shell spectrum clearly shows---accompanied by a smaller
second local maximum---a drop in the absorption
intensity on the blue side of the first main transition. Because of the limited
number of distinct realizations and the omission of effects like 
size disorder, phonon-couplings and temperature-induced broadenings, all the 
smaller peak variations will not be relevant for the comparison with an
experimental UV-Vis spectrum, which is usually of a much \quotes{smoother} shape.
Nevertheless, we think that this 
feature might be so prominent that it can also be found in an experimental absorption
spectrum for the phase-sperated CdSe/CdS system under consideration.

%%%%%%%%%%%%%%%%%%%%%%%%%%%%%%%%%
\subsection{Results for type-II (Zn,Cd)(Se,S), ZnSe/CdS and CdS/ZnSe QDs}

\begin{figure*}
\includegraphics[width=\linewidth]{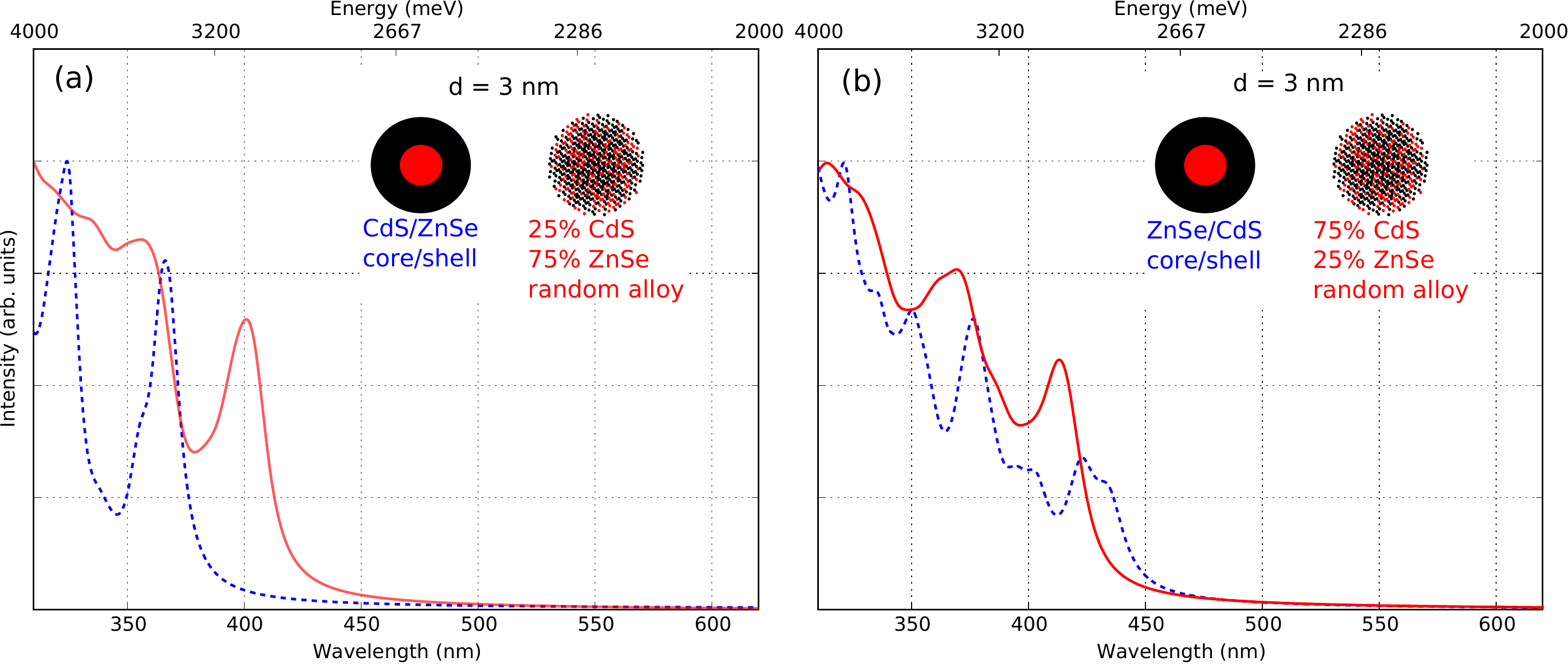}
\caption{
\label{fig:cds_znse_spectra}
\centering
Calculated absorption spectra for (a) Cd$_{0.25}$Zn$_{0.75}$S$_{0.25}$Se$_{0.75}$ randomly 
alloyed QDs (solid red) and the CdS/ZnSe core/shell counterpart (dashed blue) and 
(b) Cd$_{0.75}$Zn$_{0.25}$S$_{0.75}$Se$_{0.25}$ randomly alloyed QDs (solid red)
and their ZnSe/CdS core/shell counterpart (dashed blue).}
\end{figure*}

The absorption spectra for the
 Cd$_{0.25}$Zn$_{0.75}$S$_{0.25}$Se$_{0.75}$ randomly 
alloyed QDs and their CdS/ZnSe core/shell counterpart, as well as for 
Cd$_{0.75}$Zn$_{0.25}$S$_{0.75}$Se$_{0.25}$ randomly alloyed QDs 
and ZnSe/CdS core/shell can be found in 
Figs.\,\ref{fig:cds_znse_spectra}a and \ref{fig:cds_znse_spectra}b, respectively.

\textbf{Cd$_{0.25}$Zn$_{0.75}$S$_{0.25}$Se$_{0.75}$ vs.~CdS/ZnSe.}
We will start with a discussion of the former system, of
which the spectrum is depicted in Fig.\,\ref{fig:cds_znse_spectra}a. In contrast
to the type-I Cd(Se,S) system (where the lines were redshifted after
phase separation), we now find for the lowest alloy
absorption peak at ca.~3.10 \eV a blueshift of 290 \meV (34 nm) upon
 separation into the CdS/ZnSe core/shell system (see also  
Fig.\,\ref{fig:peakShifts}). Furthermore, the core/shell spectrum exhibits
a different shape with an emphasized intensity minimum between a prominent
first and second main 
absorption band, while the alloy spectrum has very little structure 
(a quasi-continuous absorption onset)
on the blue high-energy/low wavelength 
side of the main absorption band. According to our simulations, 
both effects (blueshift and second main absorption band) should be 
prominent enough to also be identifiable in a real experimental UV-Vis setup.

\textbf{Cd$_{0.75}$Zn$_{0.25}$S$_{0.75}$Se$_{0.25}$ vs.~ZnSe/CdS.}
We will close with a discussion of the counterpart of the 
preceding type-II system. The absorption spectrum for
Cd$_{0.75}$Zn$_{0.25}$S$_{0.75}$Se$_{0.25}$ randomly alloyed QDs 
and ZnSe/CdS core/shell is  depicted in Fig.\,\ref{fig:cds_znse_spectra}b.
While in the preceding
two cases, the random alloy and core/shell cases showed clearly different
features, this is much less the case now. The lowest alloy absorption peak at 
ca.~3.00 \eV
(see again Fig.\,\ref{fig:peakShifts}) now shows a relatively small redshift of
120 \meV (18 nm) upon phase separation. Also, the overall shape of the spectrum is slightly
different, albeit both show local minima that are seperated by the more or less 
constant small redshift. Nevertheless, the differences are so small that 
corresponding predictions will  presumably exceed the range of validity of our models, and 
would most likely not be identifiable in real UV-Vis spectra due to the already discussed additional
effects and uncertainties. Still, the general trend that we find here might be useful in further studies
on this colloidal QD system. 

%%%%%%%%%%%%%%%%%%%%%%%%%%%%%%%%%
\subsection{Further discussion}

It should be emphasized that the energetic shifts observed here are not of the
same nature as the commonly observed redshifts in many type-II structures that are induced
upon shell growth.\cite{smith_semiconductor_2010, tyrakowski_bright_2015} The latter shifts
can, e.g., result from a reduction of the effective band gap, a change in the work function
or, additionally, be strain-induced. Still, they
always refer to the bare core system with smaller diameter, 
while our reference systems are random alloys of the
same overall QD size. A random alloy QD often exhibits a much smaller gap than
the linearly averaged (\quotes{virtual crystal}) gaps of the constituents due to the
considerably large
nonlinear bowing coefficient.~\cite{mourad_multiband_2010, mourad_random-alloying_2014}
Depending on the materials and sizes under consideration, the redshift
that results from an admixture of small band gap material can
 be larger than shifts upon shell growth and/or size variations. However, the corresponding
 analysis is beyond the scope of this paper and will be part of future work.

A further analysis in analogy to Ref.\,\onlinecite{mourad_random-alloying_2014},
(which is not repeated here for the sake of brevity) 
of the wave functions and oscillator strengths for the here
examined core/shell systems shows that the wave functions significantly
leak across the barrier over the whole QD and, e.g., in ZnSe/CdS the electron ground state
(which is "CdS shell-like") has considerable overlap with the lowest hole states 
("ZnSe core-like") and is optically active. Therefore, and as already mentioned in
Sec.\,\ref{subsec:choice}, for such small cores and thin
shells, the bulk-derived designation as type-II or quasi-type-II is here of limited
explanatory power for the QD optical properties. Also, we stress that our 
non self-consistent method also has
inherent limitations when applied to very small systems. Still, it should 
be pointed out that it is always more accurate than a corresponding multiband
effective mass/$\f{k}\cdot \f{p}$ calculation, as the latter can analytically 
be proven to be a limit case
of our TB model.~\cite{loehr_improved_1994}

%%%%%%%%%%%%%%%%%%%%%%%%%%%%%%%%%%%%%%%%%%%%%%%%%%%%%%%%%%%%%%%%%%%%%%%%%%%%%
\section{Summary\label{sec:summary}}
%%%%%%%%%%%%%%%%%%%%%%%%%%%%%%%%%%%%%%%%%%%%%%%%%%%%%%%%%%%%%%%%%%%%%%%%%%%%%

In randomly alloyed monodisperse semiconductor quantum dots, 
the relative composition of the constituents is a powerful means to continuously 
tune the absorption and emission wavelength and further technologically
relevant optoelectronic properties. However, the distinction between real solid-solution
like alloy QDs and phase-separated core/shell QDs---which are technologically useful
on their own, but undesired in some cases---is difficult and usually requires a combination
of rather sophisticated tools for structural characterization.

In this paper, which represents a considerable extension of more specialized previous
works,~\cite{mourad_multiband_2010, mourad_random-alloying_2014} we have shown that
the optical absorption spectrum of small ($3$ nm diameter) QDs can in principle be 
used to differentiate 
a random alloy phase from a stoichiometrically equivalent core/shell phase.
Our model calculations are based on conceptually simple
tight-binding calculation, are exact in the substitutional disorder degree of freedom
and inhibit local alloy effects like wave function distortions and their microscopical impact
on dipole selection rules, as well as nonlinear shifts
in the electron and hole quasiparticle spectra.

Our comprehensive scheme only uses bulk band
structure properties without additional free parameters and has been applied to calculate
the random alloy vs.~core/shell absorption spectrum of
one type-I QD system, (i) CdSe$_{0.25}$S$_{0.75}$ vs.~CdSe/CdS, and two similar
type-II QD systems, (ii) Cd$_{0.25}$Zn$_{0.75}$S$_{0.25}$Se$_{0.75}$ vs.~CdS/ZnSe and 
(iii) Cd$_{0.75}$Zn$_{0.25}$S$_{0.75}$Se$_{0.25}$ vs.~ZnSe/CdS.
In particular, we 
have analysed the energetic shift of the lowest absorption 
peak upon phase change and the overall shape of the low energy/large wavelength
part of the spectrum.
For cases (i) and (ii) we find strong signatures upon core/shell desegregation that 
should also be identifiable in realistic experimental setups.
For case (iii), we also find quantifiable fingerprints, but think that they
might be too weak when compared to the overall accuracy of our method.

To give an outlook, we think that the method employed here can be a valuable tool
to circumvent more sophisticated structural analysis method for a multitude of further
systems. In the future, it can be applied to further colloidal QD systems, as well as larger
diameters. When additional disorder degrees of freedom, e.g., size and shape disorder
are applied (as long as the computational effort allows to do so), 
we expect even more accurate
predictions.

%%%%%%%%%%%%%%%%%%%%%%%%%%%%%%%%%%%%%%%%%%%%%%%%%%%%%%%%%%%%%%%%%%%%%%%%%%%%%
\begin{acknowledgments}
The author would first and foremost like to thank Z.\,Hens, but also
A.\,Guille, T.\,Aubert and
E.\,Brainis  for the fruitful collaboration and stimulating 
discussions that lead to this work.
Furthermore, the author would like to acknowledge G.\,Czycholl, R.\,Binder and N.-H.\,Kwong
for interesting discussions about the intricacies of the light-matter coupling,
is indebted to T.\,Wehling for the opportunity to carry out
this research as part of his group and thanks M.\,Sch\"uler
and J.\,Jackson for helpful comments 
on the manuscript.
Also, grants of the North-German Supercomputing Alliance
HLRN for computation time and support from the Deutsche
Forschungsgemeinschaft, Project No.\,CZ 31/20-1,
``Many-body theory of optical properties for semiconductor nanostructures 
based on atomistic tight-binding models'' are thankfully acknowledged.
\end{acknowledgments}

\begin{appendix}

\section{Dielectric mismatch effects in core/shell and alloyed quantum dots
\label{sec:Appendix}}
%
%\begin{enumerate}
% \item

\textbf{Influence of dielectric mismatch.}
In aqueous solutions $\varepsilon_\text{out}$ will be one 
 order of magnitude larger than $ \varepsilon_\text{in}$.
As $\Sigma_\text{pol}$ is predominantly a function of
$\varepsilon_\text{out}^{-1} -
\varepsilon_\text{in}^{-1}$,\cite{lannoo_screening_1995} the
magnitude of the surface-induced
polarization will then be dominated by
QD-specific variations of $\varepsilon_\text{in}^{-1}$.
Whether or not a properly defined $E^\text{coul}$ will then be subject to the 
same cancellation effect is a rather involved question (and cannot be answered
without a self-consistent approach), but of minor importance
for our present paper: All three binary semiconductor 
materials CdS, CdSe and ZnSe 
%QDs (alloyed and core/shell)
would lead to  a more or less identical site-averaged
screening coefficient
$\varepsilon_\text{in}^{-1}\approx 0.2(2)$
%$\varepsilon_\text{in}^{-1}\approx 0.22$
(see
Refs.\,\onlinecite{mourad_multiband_2010, 
mourad_random-alloying_2014} and 
references therein\cite{penn_wave-number-dependent_1962, franceschetti_many-body_1999})
for a pure binary QD of the here examined size.
Therefore, any randomly alloyed or core-shell seperated
QD can approximately be characterized by the same value 
$\varepsilon_\text{in}^{-1}$, which results in an identical
$\Sigma_\text{pol}$ for a given diameter.

Moreover, 
$\Sigma_\text{pol}$
and $E^\text{coul}$ can still be expected to approximately cancel for a 
given microstate, geometry and composition,
as---in the simplest first order approximation---both
stem from a site average weighted 
with the same probability density $|\psi(\f{r})|^2$.\cite{lannoo_screening_1995}
Within a multiband effective-mass based approach,
it 
has been shown in Ref.\,\onlinecite{fonoberov_exciton_2002} 
that the experimentally accessible excitonic gap is only weakly dependent on the 
magnitude of the dielectric mismatch at the boundary.
%
%\item

\textbf{Transferability to alloy and core/shell systems.}
Due to the above discussed almost constant $\varepsilon_\text{in}^{-1}$ for all 
materials under consideration, we assume the influence of image charge contributions
at the core/shell interface to be negligible.\footnote{A quantitative assessment
of this assumption is beyond the scope of our present model}
To see whether the cancellation assumption is still valid when comparing 
the alloyed and core/shell realizations, we performed test calculations 
of density-density like Coulomb MEs\footnote{Behold that this nomenclature refers to the electron-hole
picture,\cite{onida_electronic_2002} while in the electron-electron
picture these MEs come from an exchange-like self-energy variation.}
\begin{equation}\label{eq:CoulombMEs}
V^{ehhe}_{ijji} = \langle\psi^{e}_i |\langle\psi^{h}_j|V_\text{Coul}|
\psi^{h}_j \rangle | \psi^{e}_i \rangle,
\end{equation}
where $V_\text{Coul}$
is the Coulomb operator,
 for several electron and hole 
states and randomly selected single alloy microstates, alongside the core/shell QDs.
These MEs constitute the largest part of the the quasi-electron quasi-hole attraction.
Overall, the discrepancy in the screened Coulomb MEs between the randomly alloyed and
corresponding
core/shell QDs is roughly between 0.02 eV
(Zn$_{0.75}$Cd$_{0.25}$Se$_{0.75}$S$_{0.25}$ vs equivalent CdS/ZnSe core/shell)
and 0.05 eV (CdSe$_{0.25}$S$_{0.75}$ vs equivalent CdSe/CdS core/shell),
while we estimate the absolute 
accuracy of the absorption spectrum calculated in an empirical
TB method to 0.1 eV (at most)
in this size regime.
%
%\item

\textbf{Transferability to higher transitions.}
Although the results from Refs.\,\onlinecite{lannoo_screening_1995,
delerue_excitonic_2000, fonoberov_exciton_2002} were only obtained for 
the energetically
lowest transitions of unalloyed systems,
%(eventually dipole-forbidden ground state transitions can practically 
%be neglected in the discussion, as the hole levels lie very dense) 
we found 
the basic conclusions also to be approximately valid 
for a large part of the spectrum of bound states
of disordered systems in the energetic vicinity of the excitonic gap 
(see the discussion in Ref.\,\onlinecite{mourad_random-alloying_2014}). This
can presumably be traced back to the fact that we use an empirical TB model 
that does not only
reliably fit the Brillouin zone center, but the dispersion over the whole
irreducible wedge for the optically relevant bands,~\cite{loehr_improved_1994,
marquardt_comparison_2008} 
which corresponds to
$\delta \Sigma_\text{bulk}\approx 0$ over the whole 
first Brillouin zone. While $GW$ calculations 
practically yield  wave-vector independent self-energy corrections 
to the conduction bands for
most semiconductors like Si and GaAs~,\cite{godby_self-energy_1988} 
the proper modeling can be more complicated for the strongly ionic
II-VI compounds examined here for a multitude of
reasons.~\cite{zakharov_quasiparticle_1994,
rohlfing_quasiparticle_1995} Within our empirical TB model, we are not dependent
on the validity of any of those assumptions and their impact on the 
calculation of higher transitions in II-VI quantum dots
and---if available---can also incoporate any additional corrections based
on experimental findings.
On the other hand, the here employed rather artificial 
assumption of perfect surface passivation will have a larger impact 
on higher excited states,
so we expect our model to be less accurate when the absorption
energies approach the experimentally known ionization continuum.

\textbf{Impact of disorder effects to Coulomb matrix elements.}
As already discussed in Refs.\,\onlinecite{mourad_multiband_2010, 
mourad_random-alloying_2014},
the most prominent consequences of the additional disorder degree of freedom are
(i) the resulting large scattering of the single-particle energy levels and
(ii) the presence of additional dipole transitions upon the alloy-induced 
wave function distortion. The disorder-related scattering of the Coulomb MEs 
takes place on a scale of tens of \meV for the sizes
under consideration (see Fig.\,6 of Ref.\,\onlinecite{mourad_multiband_2010}) 
and is thus of minor importance
for the discussion of the visible spectrum on a corresponding energy scale.

\end{appendix}
%%%%%%%%%%%%%%%%%%%%%%%%%%%%%%%%%%%%%%%%%%%%%%%%%%%%%%%%%%%%%%%%%%%%%%%%%%%%%

%merlin.mbs aipnum4-1.bst 2010-07-25 4.21a (PWD, AO, DPC) hacked
%Control: key (0)
%Control: author (8) initials jnrlst
%Control: editor formatted (1) identically to author
%Control: production of article title (-1) disabled
%Control: page (0) single
%Control: year (1) truncated
%Control: production of eprint (0) enabled
%

\end{document}